\documentclass{emulateapj}
\usepackage{epsfig}

\newcommand{\gx}{GX\,339--4}

\newcommand{\nad}{Na~D}
\newcommand{\NaD}{Na~D}

\newcommand{\HeII}{He\,{\sc ii}}
\newcommand{\NaI}{Na\,{\sc i}}
\newcommand{\CaII}{Ca\,{\sc ii}}

\newcommand{\ebv}{E(B-V)}

\slugcomment{Accepted for publication in the Astrophysical Journal,
  2004 March 10}

\shorttitle{The Distance to GX\,339--4}
\shortauthors{Hynes et al.}

\begin{document}


\title{The Distance and Interstellar Sight Line to GX\,339--4}

\author{R. I. Hynes\altaffilmark{1,2},
        D. Steeghs\altaffilmark{3},
        J. Casares\altaffilmark{4}, 
	P. A. Charles\altaffilmark{5},
	and
	K. O'Brien\altaffilmark{6}}
\altaffiltext{1}{McDonald Observatory and Astronomy Department, The University of Texas at Austin, 
1 University Station C1400, Austin, Texas 78712, USA; rih@astro.as.utexas.edu}
\altaffiltext{2}{Hubble Fellow}
\altaffiltext{3}{Harvard-Smithsonian Center for Astrophysics, 
60 Garden Street, MS-67, Cambridge, MA 02138, USA; 
dsteeghs@head-cfa.harvard.edu}
\altaffiltext{4}{Instituto de Astrof\'\i{}sica de Canarias, 38200 La
  Laguna, Tenerife, Spain; jcv@ll.iac.es}
\altaffiltext{5}{School of Physics and Astronomy, 
The University of Southampton, Southampton, SO17 1BJ, UK; pac@astro.soton.ac.uk}
\altaffiltext{6}{European Southern Observatory, Casilla 19001,
        Santiago 19, Chile; kobrien@eso.org}

\begin{abstract}

The distance to the black hole binary \gx\ remains a topic of debate.
We examine high-resolution optical spectra of the \nad\ lines
resolving the velocity structure along the line of sight.  We find
this to be complex, with at least nine components, mostly
blue-shifted, spanning a velocity range of nearly 200\,km\,s$^{-1}$.
The presence of components with a large blue-shift rules out a nearby
location and requires that the binary be located at or beyond the
tangent point, implying a lower limit to the distance of $\sim6$\,kpc.
The presence of a significant red-shifted component at
$+30$\,km\,s$^{-1}$ is even more intriguing as \gx\ also has a
slightly positive systemic velocity, suggesting that the source, and
this cloud, could be on the far side of the Galaxy, where the radial
velocities due to Galactic rotation become positive again.  If this is
the case, we require a distance of $\sim15$\,kpc.  This is less
secure than the 6\,kpc lower limit however.  We discuss the
implications of these possible distances for the outburst and
quiescent luminosities, and the nature of the companion star, and
argue that a large distance is consistent with these
characteristics.  In particular, it would explain the non-detection of
the companion star during the faintest states.  

\end{abstract}

\keywords{stars: individual: GX\,339--4}

\section{Introduction}

The distance of X-ray binaries is challenging to determine precisely.
They are usually too distant for current parallax measurements, and
there are no really reliable standard candles, other than type-I X-ray
bursts showing radius expansion, and these only occur in a subset of
neutron star systems (see \citealt{Kuulkers:2003a} and references
therein).  For black holes, such as \gx, the problem is much less
tractable.  Fortunately the majority of such systems undergo periods
of deep quiescence when it is possible to detect the companion star,
and determine the binary parameters.  Hence both the radius and
temperature (i.e.\ surface brightness) can be estimated and the
distance derived from the apparent brightness (see e.g.\
\citealt{Gelino:2001a}).  For \gx\ this is not possible, as even in a
very low luminosity state, no spectral features of the companion star
were seen in VLT/FORS observations \citep{Shahbaz:2001a}.  This is
frustrating for distance and parameter estimation, and very
puzzling in its own right.  One explanation for the non-detection of
the companion is that the distance has been significantly
underestimated.

Current distance estimates are very uncertain; \citet{Buxton:2003a}
has provided a convenient summary.  Early estimates were rather low,
mostly falling in the 1--4\,kpc range.  More recently, larger
estimates have been proposed.  The non-detection of the companion star
\citep{Shahbaz:2001a} was used as evidence for a distance $\ga
5.6$\,kpc.  A more novel approach was taken by \citet{Maccarone:2003a}
who has shown that in most black hole binaries, transitions between
the high/soft and low/hard states occur at approximately a fixed
fraction of the Eddington limit.  It was suggested that the transition
luminosity can then be used as a mass-dependent standard candle.
Assuming a minimum black hole mass for \gx\ of $\ga5.8$\,M$_{\odot}$
\citep{Hynes:2003a}, \citet{Maccarone:2003a} estimate a distance $\ga
6$\,kpc, allowing for scatter about the relation.

The actual distance of \gx\ thus remains very uncertain, but could be
rather large.  This introduces considerable uncertainty into
interpreting its quiescent and outburst luminosities, and the nature
of the companion.  The distance also has important implications for
interpretation of the recently discovered resolved jets;
\citet{Gallo:2004a} estimated that the motion of the jet head requires
an apparent velocity of 0.9\,c even at 4\,kpc.  A significantly larger
distance would therefore imply apparent superluminal motion.  We have
therefore analyzed interstellar features in high resolution spectra
described in Section~\ref{ObservationSection} to elucidate the
properties of the line-of-sight in more detail.  In
Section~\ref{ModelSection} we perform a detailed study of the
properties of line-of-sight \NaI\ and \CaII\ absorption, resolving
most individual clouds in \NaI\ and hence determining cloud-by-cloud
velocities, velocity dispersions, and \NaI\ column densities.  We also
have some information about the relative abundances of \CaII\ to \NaI\
as a function of velocity.  We discuss the implications for the
distance of \gx\ in Section~\ref{RotationSection}, and for the
reddening in Section~\ref{ReddeningSection}.  Finally in
Section~\ref{DiscussionSection} we consider how the range of distances
proposed affect estimates of the outburst and quiescent X-ray
luminosities and the nature of the companion star, and in
Section~\ref{ConclusionSection} we summarize our conclusions.
%
%
\newpage

\section{Observations}
\label{ObservationSection}
High-resolution observations of \gx\ were obtained with the UV-Visual
Echelle Spectrograph (UVES) on UT2 (Kueyen) at the VLT on 2002 August
9--15.  These spanned a wavelength range of 4100--6200\,\AA\ and
included a total exposure time of 12\,hr yielding a superb quality
average spectrum (signal-to-noise ratio near the \nad\ lines greater
than 100 without binning).  A 1'' slit was used giving a nominal
resolution near the \nad\ lines of $\simeq44\,000$, corresponding to a
kinematic resolution of 6.8\,km\,s$^{-1}$.  The starlight will be
concentrated to the center of the slit, however, yielding a somewhat
higher resolution.  From the spatial profiles near \nad\ we estimate a
mean seeing of about 0.5'', implying that the resolution could be as
good as 4\,km\,s$^{-1}$ with a 1'' slit.  In practice, this will not
be realized and we expect a value between these extremes.  We
therefore leave the instrumental resolution as a free parameter in
subsequent fits.  For the rest of this work we will focus only upon
the \nad1 and D2 lines (5895.92\,\AA\ and 5889.95\,\AA\ respectively).
Pipeline optimal extractions were supplied and were of good quality in
this region; a manual reduction of selected spectra yielded negligible
improvement.

We supplement these data with lower-resolution observations of the
\CaII\ H and K lines (3968.47\,\AA\ and 3933.66\,\AA\ respectively).
These were obtained with the RGO spectrograph of the Anglo-Australian
Telescope over 2002 June 6--11, with the R1200B grating yielding a
wavelength coverage of 3500--5250\,\AA.  The images were de-biased and
flat-fielded, and the spectra subsequently extracted using
conventional optimal extraction techniques \citep{Horne:1986a}.
Wavelength calibrations were interpolated between CuAr comparison lamp
images, obtained every 20--30\,mins.  The kinematic resolution is
estimated to be 74\,km\,s$^{-1}$.

%
\section{Line-of-sight velocity structure}
\label{ModelSection}
\subsection{\NaI\ absorption}
\begin{figure}
\epsfig{angle=90,width=3.0in,file=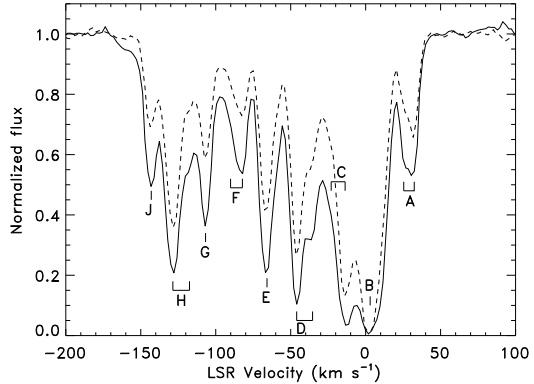}
\caption{Line profiles of the \nad1 (dashed) and \nad2 (solid) lines.
  Spectra have been normalized, but no offset is applied between the
  lines.  Labels denote distinct components.  Some of these (e.g.\ D)
  appear to be further unresolved multiples.  Tick marks indicate the
  components fitted in Table~\ref{OptFitTable}.}
\label{NaDFig}
\end{figure}

We show line profiles of the two \nad\ lines in Fig.~\ref{NaDFig}
relative to the local standard of rest (LSR).  They are clearly
complex with a large velocity dispersion, suggesting a relatively
large distance.  This immediately rules out suggestions that \gx\ is a
local object at $\sim 1.3$\,kpc (\citealt{Mauche:1986a};
\citealt{Predehl:1991a}); the same conclusion was drawn by
\citet{Zdziarski:1998a} for different reasons.

\begin{figure}
\epsfig{angle=90,width=3.0in,file=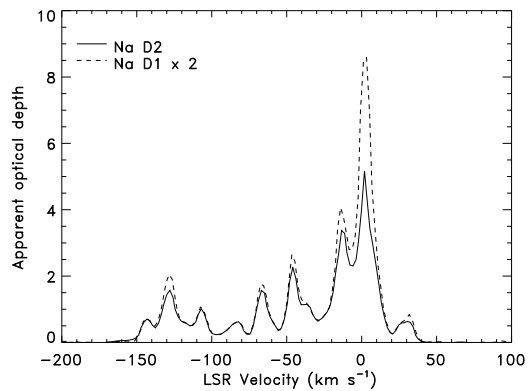}
\caption{Apparent optical depths of the \nad\ lines.  The D2 optical
  depths have been doubled so that they are identical to D1 in the
  optically thin limit.  Where the lines disagree, saturation is
  important and the optical depth may be underestimated.}
\label{NaDTauFig}
\end{figure}

We can obtain an approximate estimate of the distribution of \NaI\
with velocity by converting the line profiles into apparent optical
depths (Fig.~\ref{NaDTauFig}), as discussed by \citet{Sembach:1993a}.
This is not the true optical depth, because it is modified by
instrumental broadening. It is also reduced (relative to the \NaI\
column density) for strongly saturated components.  This is seen in
differences in the optical depths inferred from the D1 and D2 lines.
We have scaled-up the D1 optical depths in Fig.~\ref{NaDTauFig} by a
factor of two (the ratio of oscillator strengths) so that the two
lines can be directly compared.  Where the two optical depth profiles
overlay (i.e.\ the D1 line has exactly half the optical depth of the
D2 line) the apparent optical depths are simply blurred versions of
the true optical depth, and the relative contributions of different
components are preserved.  Where unresolved saturated structure is
present, the D2 line becomes preferentially more saturated, the D1
line becomes proportionately stronger, and the ratio of apparent
optical depths approaches unity.  In this case the total absorption of
a component is underestimated, but the D1 optical depths reflect the
column densities more accurately.  We can see that most of the
absorption column appears to be at low velocities, $\left|V_{\rm
LSR}\right| \la 20$\,km\,s$^{-1}$, especially when one considers that
these are the most saturated components, and hence the most severely
underestimated.  There is also a significant contribution at higher
velocities, however.

We can perform a more quantitative analysis by modeling the line
profiles, and explicitly accounting for unresolved saturation.  We
follow standard techniques (e.g.\ \citealt{Spitzer:1978a};
\citealt{Sembach:1993a}), as also applied by \citet{Dubus:2001a} to
\CaII\ lines in XTE~J1118+480.  We assume that the line-of-sight
absorption consists of a number of discrete clouds, each of which has
mean velocity $v$, a Gaussian distribution of velocities with width
$b$, and \NaI\ column density $n$.  Given this model we can
simultaneously predict both D1 and D2 intrinsic line profiles.  We
then convolve these with the instrumental resolution and compare with
the data.  As discussed above, the exact instrumental resolution is
uncertain and was fitted as a free parameter.  The best fit was
obtained for 6.0\,km\,s$^{-1}$, which falls within the range expected
given the slit width and seeing (Section~\ref{ObservationSection}).
Note that in most cases these parameters are relatively well
constrained, even though the intrinsic widths are not fully resolved.
This is because the differential saturation of the D1 and D2 lines
makes the ratio of the two line strengths sensitive to the intrinsic
widths.  Hence we can recover a reasonable estimate of the column
density even for somewhat saturated lines.  The largest uncertainty
exists where significant saturation is present, and then we sometimes
only constrain a lower limit upon the column density.  Fitting was
performed iteratively using the Levenberg-Marquardt algorithm (see
e.g.\ \citealt{Press:1988a}).  We began by fitting the components
labeled in Fig.~\ref{NaDFig}.  Additional components were added when
it was clear that they were needed; for example D clearly includes an
additional low velocity component and a weak component is also needed
between G and H.  Other deficiencies were apparent only by comparison
with the preliminary fits and lead us to add additional components
near C and F as well.  It may be that such weak components from small
clouds are actually present throughout the profile, but are only
discernible in the gaps between the larger components (Redfield priv.\
comm.)

The derived parameters for the components are listed in
Table~\ref{OptFitTable}.  This model reproduced the data extremely
well, with no significant residuals, so we have not shown the fit.
Velocities are derived from both D1 and D2 lines simultaneously, and
indicate a range of mean cloud velocities of --143 to
+32\,km\,s$^{-1}$.  We will compare this to Galactic rotation curves
in Section~\ref{RotationSection}.  Component widths are
1.5--7.3\,km\,s$^{-1}$.  These values are typical of the ISM;
\citet{Sembach:1994a} found average values for the \nad\ lines of
4.0\,km\,s$^{-1}$.  If the lowest widths are purely thermal, they
indicate cloud temperatures of $\ga3\,000$\,K, but turbulent
broadening could be present as well.  It may be that some of the
larger widths, such as for component B, indicate additional unresolved
components rather than intrinsically high widths.  One of these
mechanisms is needed for the highest widths, as these are too high for
thermal broadening alone.  We derive a total \NaI\ column density of
at least $3.5\times10^{13}$\,cm$^{-2}$, of which components B and C
(assumed to be relatively local) contribute at least 63\,\%.

\begin{table}
\caption{Results of fits to the \nad\ lines.}
\label{OptFitTable}
\begin{tabular}{lccc}
\hline
\noalign{\smallskip}
Component & $v$\tablenotemark{a} & $b$\tablenotemark{b} & $n$\tablenotemark{c}\\
          & (km\,s$^{-1}$) & (km\,s$^{-1}$) & ($10^{12}$\,cm$^{-2}$) \\
\noalign{\smallskip}
\hline
\noalign{\smallskip}
\multicolumn{4}{l}{Single component B} \\
A1 & $  32.4^{+0.6}_{-0.3}$ & $1.5^{+0.2}_{-0.5}$ &  $0.80^{+0.15}_{-0.23}$ \\
A2 & $  25.4^{+1.2}_{-0.4}$ & $2.2^{+2.3}_{-0.5}$ &  $0.45^{+0.05}_{-0.04}$ \\
B  & $   3.0\pm0.1$         & $7.3\pm0.2$         & $12.7^{+0.7}_{-0.5}$ \\
C1 & $ -13.7^{+0.2}_{-0.3}$ & $2.7^{+0.5}_{-2.7}$ &  $8.3^{+\infty}_{-2.6}$ \\
C2 & $ -23.0^{+1.8}_{-1.6}$ & $7.4^{+1.9}_{-2.4}$ &  $1.17\pm0.35$ \\
D1 & $ -35.4\pm0.3$         & $2.2^{+0.5}_{-0.3}$ &  $0.94^{+0.10}_{-0.09}$ \\
D2 & $ -46.0\pm0.1$         & $4.0\pm0.2$         &  $3.05^{+0.14}_{-0.12}$ \\
E  & $ -65.8\pm0.1$         & $5.1\pm0.2$         &  $2.04\pm0.05$ \\
F1 & $ -82.2^{+0.3}_{-0.4}$ & $2.1^{+0.6}_{-0.4}$ &  $0.53\pm0.08$ \\
F2 & $ -90.1^{+1.0}_{-1.2}$ & $6.6^{+1.2}_{-1.3}$ &  $0.45^{+0.07}_{-0.08}$ \\
G  & $-106.8\pm0.2$         & $3.7^{+0.4}_{-0.3}$ &  $1.10\pm0.04$ \\
H1 & $-117.5\pm0.4$         & $1.2^{+0.6}_{-0.4}$ &  $0.44^{0.42}_{-0.05}$ \\
H2 & $-128.6\pm0.1$         & $4.4\pm0.3$         &  $2.32^{+0.09}_{-0.08}$ \\
J  & $-143.0\pm0.2$         & $2.7^{+0.4}_{-0.3}$ &  $0.75^{+0.06}_{-0.04}$ \\
\hline
\multicolumn{4}{l}{Split component B\tablenotemark{d}} \\
B1 & $  12.6^{+0.5}_{-0.7}$ & $3.1^{+0.7}_{-0.5}$ & $  1.06^{+0.21}_{-0.13}$   \\ 
B2 & $   2.3\pm0.1$         & $3.0^{+0.4}_{-0.5}$ &   $250^{+1060}_{-150}$ \\
C1\tablenotemark{e}
 & $ -13.3\pm0.2$           & $5.5^{+0.4}_{-0.5}$ & $  5.20^{+0.16}_{-0.15}$   \\
C2 & $ -26.4^{+0.6}_{-0.5}$ & $3.1^{+1.4}_{-1.1}$ & $  0.53^{+0.12}_{-0.07}$   \\
\hline
\end{tabular}
\tablenotetext{a}{Mean velocity.}
\tablenotetext{b}{Gaussian width.}
\tablenotetext{c}{\NaI\ column density.}
\tablenotetext{d}{Revised parameters for B and C.  Other components
  are unaffected.}
\tablenotetext{e}{There is no formal lower limit on the width of 
component C1, or a corresponding upper limit on the column density.}
\end{table}

In most cases the sensitivity to unresolved blends is weak as the
components are not strongly saturated and the total column density is
approximately conserved.  For component B, however, the total column
density is sensitive to assumed substructure.  To illustrate this, we
repeated the fitting with two sub-components; the results are shown in
Table~\ref{OptFitTable}.  This allows the main component to be
narrower, and hence saturate more easily.  A much larger column
density is then required to achieve the same equivalent width of
absorption.  Consequently, the single component fit should be
considered the lower limit on the column density in the BC complex
($2.2\times10^{13}$\,cm$^{-2}$).  This implies that the total column
density estimated above, and the fraction of total column contained
within the BC complex, are also lower limits.  Similar difficulties
exist for components C1 and H1, and large column densities cannot be
formally rejected in either case.  This reflects the significant
saturation also present in these components (Fig.~\ref{NaDTauFig}).

\subsection{\CaII\ absorption}

\begin{figure}
\epsfig{angle=90,width=3.0in,file=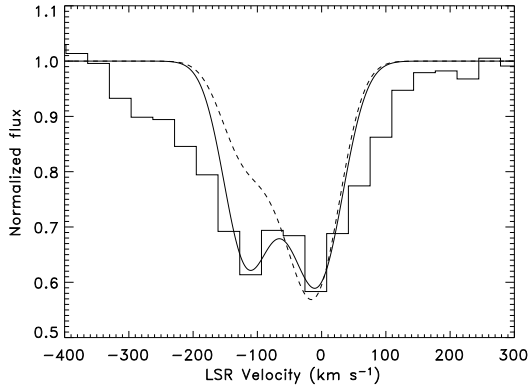}
\caption{Line profile of the \CaII\ K line.  Histograms show the data,
at much lower resolution than for the \nad\ lines.  The dashed line
shows a `blind' prediction of the line profile based on the \nad\
fits, assuming no velocity correction is needed, that the intrinsic
component widths are thermal, and that the ratio of
\CaII\ to \NaI\ is 0.6, appropriate for inner Galactic center lines of
sight \citep{Sembach:1994a}.  The solid line is a better fit including
a small velocity offset and allowing the abundance of the G--J complex
to differ from the rest of the components.  See the text for more
details.}
\label{CaIIFig}
\end{figure}

The resolution of the AAT observations of the \CaII\ lines is not
sufficient to perform a similar component-by-component analysis.  We
can, however, use the model derived from the \nad\ lines to predict
the lower-resolution profiles and perform a crude comparison.  To do
this we initially assume a ratio of \CaII\ to \NaI\ column densities
of 0.6, appropriate for inner Galactic center sight-lines
\citep{Sembach:1994a}, and an instrumental resolution of
74\,km\,s$^{-1}$.  The intrinsic width of each component is scaled
down by 0.76, which is appropriate if it is dominated by thermal
broadening (i.e.\ scaling by the square root of the atomic weight).
The fit to the profile of the K line is shown in Fig.~\ref{CaIIFig};
the H line is not used because it is blended with Balmer emission, but
the line profile otherwise appears to be similar.  The abundance ratio
assumed clearly underestimates the \CaII\ absorption at higher
velocities.  To improve the fit, we recalculated assuming a small
velocity offset (5\,km\,s$^{-1}$; a small fraction of a resolution
element) for the AAT spectra, a slightly lower ratio of \CaII\ to
\NaI\ of 0.53 for most components, but a higher ratio of 4.5 for
components G--J.  If we alternatively keep the intrinsic width, $b$,
fixed (appropriate for turbulent broadening) then we require abundance
ratios of 0.42 and 2.5 for low and high velocity components
respectively.  Either model provides a much better fit, although
clearly still underestimates the wings of the \CaII\ profiles.  This
is a common feature; additional \CaII\ absorption is often seen at
high velocities (beyond those allowed by Galactic rotation) where
\NaI\ absorption is weak or absent \citep{Sembach:1994a}.  It has been
attributed to a significantly enhanced abundance of gas-phase calcium
(by a factor of 50 or more relative to sodium) in the high velocity
components of the ISM via collisional destruction of dust grains.  It
is also observed that the mean calcium abundance is enhanced along
halo sight lines \citep{Sembach:1994a}, although this may simply
reflect the preponderance of high velocity material at high latitudes.
Other than such diagnostics, the \NaI\ lines are more useful, as they
are more closely related to Galactic rotation, and correlate better
with $\ebv$.

%
\section{The distance to \gx}
\label{RotationSection}
\begin{figure}
\epsfig{angle=90,width=3.0in,file=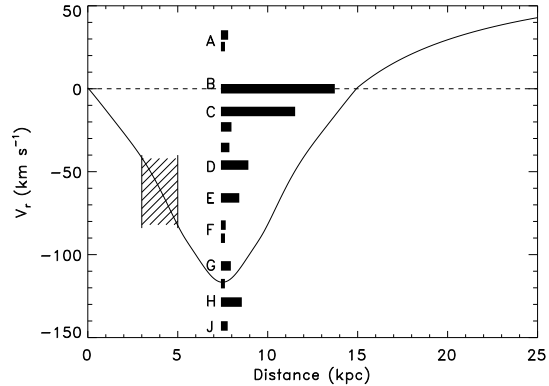}
\caption{Expected relation between radial velocity and distance along
  the line of sight to \gx.  The curve uses the rotation curve adopted
  by \citet{Nakanishi:2003a}.  Each horizontal bar represents one of
  the fitted components of \nad\ absorption.  The vertical position
  indicates the fitted velocity and the length of the bar is
  proportional to the column density.  The horizontal positions are
  arbitrary.  The hatched region indicates the distance range favored
  by \citet{Zdziarski:1998a}.}
\label{RotCurveFig}
\end{figure}

We have mapped out the properties of gas clouds along the
line-of-sight to \gx\ in velocity space.  We now attempt to associate
these with distance, assuming the velocity is due to Galactic
rotation, and hence derive a lower-limit on the distance to \gx.
Qualitatively we expect to see low-velocity material from the Solar
neighborhood, and increasingly negative velocities as the tangent
point is approached.  Beyond this, velocities become less negative
again, and become positive at distances outside the Solar circle.  The
preponderance of absorption components with negative velocities is
thus as expected from Galactic rotation along this sight line.  To be
more quantitative, we show the expected velocity as a function of
distance in Fig.~\ref{RotCurveFig}.  We assumed the rotation curve
prescription of \citet{Clemens:1985a}, with the modified polynomial
coefficients of \citet{Nakanishi:2003a}.  These were derived for a
Solar Galactocentric radius of $R_0 = 8.0$\,kpc and velocity of
$\Theta_0 = 217$\,km\,s$^{-1}$, consistent with the most recent
geometric determination of the distance to the Galactic center
($R_0=7.94\pm0.42$\,kpc, $\Theta_0 = 220.7\pm12.7$\,km\,s$^{-1}$;
\citealt{Eisenhauer:2003a}).  Fig.~\ref{RotCurveFig} also indicates
the velocity and relative column density of each fitted absorption
component for easy comparison with the predicted velocities.

Components B and C in \gx\ likely correspond to local material, and
contribute $\ga63$\,\%\ of the total \NaI\ column density.  The
velocities correspond to a distance $\la 2$\,kpc, but the material
could be within 1\,kpc allowing for a 10\,km\,s$^{-1}$ dispersion in
mean velocities.  For a Galactic latitude of $-4.5^{\circ}$, this
material will thus be $\la0.16$\,kpc below the plane and well within
the \NaI\ layer (which has an estimated scale height
$0.43^{+0.12}_{-0.08}$\,kpc; \citealt{Sembach:1994a}).  Examination of
field stars near \gx, at distances up to 2\,kpc, suggests that local
material within this distance accounts for $\sim2/3$ of the reddening
to \gx, consistent with this \citep{Zdziarski:1998a}.

If the distance to \gx\ were $\sim4$\,kpc, as proposed by
\citet{Zdziarski:1998a} and \citet{Buxton:2003a}, then the maximum
expected negative velocity would be $\sim -60$\,km\,s$^{-1}$.
Components D and E could be co-rotating, but the G--J complex would
require peculiar velocities of 40--80\,km\,s$^{-1}$.  The latter
accounts for $\la13$\,\%\ of the total column density, and as we have
already noted is relatively overabundant in \CaII.  There are two
possible interpretations for this high velocity absorption.  We
consider the most likely to be that it is co-rotating with the
Galactic rotation curve.  In support of this, the similarity of the
velocities of the G--J complex to the predicted tangent point velocity
of approximately $-120$\,km\,s$^{-1}$ is suggestive.  The range of
velocities is also in very good agreement with CO maps, which exhibit
a vertically integrated range of velocities to $-140$\,km\,s$^{-1}$
for this Galactic longitude \citep{Dame:2001a}.  The inferred altitude
of the tangent point material, is $\sim0.55$\,kpc.  This is not a
serious problem given a scale height of 0.43\,kpc.  The alternative
explanation is that the G--J complex (and component F) arises from
closer material with large peculiar velocities; the same material as
the `forbidden velocity' gas of \citet{Sembach:1994a}.  The enhanced
\CaII\ abundance would be typical of such gas.  We believe that the
presence of multiple components contributing a significant fraction of
the \NaI\ column at such high velocities is very unlikely, however.
Such features are not seen in \NaI\ in any of the similar lines of
sight collected by \citet{Sembach:1993a}, and \citet{Sembach:1994a}
estimated that even the velocity dispersion of the {\em high velocity}
\CaII\ absorption components was only $21.3$\,km\,s$^{-1}$.  If \gx\
lies at 4\,kpc it would therefore have a rather unusual line-of-sight.
Neither are such extreme velocities seen in any of the spectra of
local stars presented by \citet{Welty:2001a}.  Consequently, it seems
more likely that most of the absorption velocities do indicate
Galactic rotation and that \gx\ lies at or beyond the tangent
point. The implied distance is then $\ga 6$\,kpc, allowing for some
uncertainty in rotation curves and the intrinsic velocity dispersion
of the absorbing clouds.

Beyond this, the situation is less clear.  This is because the
velocities increase again, and so there is ambiguity between close and
distant material.  Components D--F could arise on either side of the
tangent point, and it is hard to make a conclusive statement about how
far beyond this \gx\ could lie.  Two lines of argument suggest a
significantly larger distance is possible, but neither are conclusive.
One is the systemic velocity of \gx\ itself.  The best estimate of
this comes from the emission line radial velocity analysis of the same
dataset used here \citep{Hynes:2003a}, with \HeII\ emission line wings
suggesting $\gamma=+12.3\pm1.0$\,km\,s$^{-1}$, and Bowen lines
consistent with this for the preferred orbital period.  This would of
course be expected for a local object, but the \NaD\ velocity
dispersion rules this out, even if earlier arguments do not.  If \gx\
lies at the tangent point, however, then the peculiar radial velocity
would be $\sim+140$\,km\,s$^{-1}$.  This is unusually large for a
black hole binary, as \citet{White:1996a} estimated that their radial
velocity dispersion should be 40\,km\,s$^{-1}$, based on their
distances from the plane and consistent with measurements.  Larger
values are not impossible (e.g.\ GRO~J1655--40 has an estimated
peculiar radial velocity of 114\,km\,s$^{-1}$;
\citealt{Brandt:1995a}), but are thought to arise when black holes are
formed by fallback onto a neutron star which received a kick
\citep{Fryer:2001a}.  A large velocity is only expected for a low-mass
black hole, however, which is inconsistent with the high mass function
of \gx\ and lack of evidence for a high inclination.  Further
increasing the distance is therefore appealing as it would reduce this
required peculiar velocity.  A location close to the Solar circle on
the far side of the Galaxy ($\ga15$\,kpc) would require negligible
peculiar velocity, consistent with most Galactic black hole binaries.
The other argument comes from the one absorption feature we have not
yet accounted for: component A.  This has two sub-components with {\em
positive} velocity, 25--35\,km\,s$^{-1}$.  If this arises locally then
it arises from material with a forbidden velocity
$\ga25$\,km\,s$^{-1}$.  Given that there are only two weak components
(or one with a complex profile), and that the total contribution to
the column density is $\la4$\,\%, this is not impossible.  This
material could also arise on the far side of the Galaxy, and would
then have a radial velocity close to that of the source.  One could
argue that since \gx\ has a Galactic latitude of $-4.5^{\circ}$, at
this distance, the line-of-sight should be more than 1\,kpc below the
plane.  However the Galaxy does exhibit warps, and the outer disk
warps downward toward the line-of-sight of \gx\
\citep{Nakanishi:2003a}.  It is possible that \gx, and the red-shifted
absorption, are associated with this warped material.

We thus conclude that a minimum distance of 6\,kpc is required, based
on the presence of intervening material with large blue-shifts,
consistent with expected tangent point velocities.  The presence of
red-shifted absorption, together with the low systemic velocity of
\gx\ itself, suggest that a much larger distance $\ga15$\,kpc might be
correct, although a large peculiar velocity for \gx, together with
forbidden velocity gas could obviate this.  \gx\ is thus further away
than most previous estimates (but see \citealt{Shahbaz:2001a} and
\citealt{Maccarone:2003a}), and possibly a lot further.
%
%
\section{The reddening of \gx}
\label{ReddeningSection}
Our high quality spectra should also allow us to obtain more robust reddening
constraints from the \nad\ lines; we do not attempt this with the \CaII\
lines as these do not trace the dust as well \citep{Sembach:1994a}.

The integrated \nad\ equivalent width (EW) has been estimated by a
number of authors.  The most recent values are 3.7\,\AA\
\citep{Soria:1999a} and $4.2\pm1.0$\,\AA\ \citep{Buxton:2003a}.  We
estimate a total EW of $3.57\pm0.02$\,\AA\ from our resolved \nad\
lines, with $1.48\pm0.01$\,\AA\ and $2.10\pm0.01$\,\AA\ in the \nad1
and D2 lines respectively, consistent with other estimates.  The
uncertainties quoted are conservative estimates based on the range of
possible continuum levels; formal statistical errors are smaller than
this.  The interpretation of the equivalent width is obviously
complicated by saturation of some components, but neglecting this we
recover a lower-limit of $\ebv\ga0.6$ from the \nad2 line assuming the
asymptotic linear calibration of \citet{Munari:1997a}.  We can perform
an analogous calculation using the less saturated \nad1 line, since in
the optically thin limit the D2/D1 ratio has a value of 2.  This
yields a lower limit of $\ebv\ga0.85$.

An alternative approach is to circumvent the saturation effects by
using the column density derived from the line profile fits.  As with
all such calibrations, it is still sensitive to assumptions about
gas-to-dust scaling ratios.  We use the values derived by
\citet{Sembach:1994a}, which include scalings for a number of types of
sight line.  An \NaI\ column of at least $3.5\times10^{13}$\,cm$^{-2}$
corresponds to $N_{\rm H} = 3.1\times10^{21}$\,cm$^{-2}$ and $\ebv =
0.50$ using ratios appropriate to inner Galactic center sight lines.
These values change to $N_{\rm H} = 3.3\times10^{21}$\,cm$^{-2}$ and
$\ebv=0.69$ using Galactic center ratios, and $N_{\rm H} =
4.4\times10^{21}$\,cm$^{-2}$ and $\ebv=0.59$ for averaged ratios.  The
uncertainty is larger than implied by the scatter between these
estimates, however, since the variance in ratios within each of these
samples is large.  While some of the variance may simply reflect
measurement errors, it is likely that the ratios are subject to
significant cloud-to-cloud scatter.  As discussed earlier, these are
also lower limits, since the local components of \nad\ are saturated
and the column density used here may have been significantly
underestimated.  This is likely why both the $N_{\rm H}$ and $\ebv$
values we estimate are lower than other measurements (e.g.\
\citealt{Zdziarski:1998a}; \citealt{Corongiu:2003a};
\citealt{Miller:2003a}).

Consequently, even with the detailed resolution presented here, we
cannot determine a reliable $\ebv$ from the \nad\ lines; this method
is fundamentally rather crude, and limited by saturation.  All we can
say with confidence is that the reddening is greater than 0.5, and
probably greater than about 0.8.  This is comparable to previous
estimates including $1.2\pm0.1$ \citep{Zdziarski:1998a} and
$1.1\pm0.2$ \citep{Buxton:2003a}.  We suspect that these quoted
uncertainties are underestimates, however, as similar difficulties
apply as with our own measurements; indeed the latter estimate is also
based on \NaD.  Upper limits are problematic, and neither the methods
of \citet{Zdziarski:1998a}, nor \citet{Buxton:2003a}, nor our
discussion above adequately address this.  Neutral hydrogen column
densities do allow a crude upper limit, however.  A range of values
are derived (see references above), but all imply $N_{\rm H} \la
7\times10^{21}$\,cm$^{-2}$.  Similarly, a range of gas-to-dust
scalings exist, but that of \citet{Bohlin:1978a}, with an assumed
30\,\%\ uncertainty, encompasses all of them, and also accounts for
the variance between lines of sight.  Using this scaling and the upper
limit on $N_{\rm H}$ implies $\ebv \la 1.6$.  The actual value could
well be much less than this.  An independent check is provided by the
Galactic dust maps of \citet{Schlegel:1998a} which imply
$E(B-V)\sim0.9$, although their use for Galactic latitudes less than
$5^{\circ}$ is not recommended by the authors.

%
%
\section{Discussion}
\label{DiscussionSection}
\subsection{The outburst luminosity}
The luminosity of the transition between high/soft and low/hard states
in black holes is of interest \citep{Maccarone:2003a} and it may even
be an approximate standard candle.  As already noted,
\citet{Maccarone:2003a} estimated a minimum distance of 6\,kpc
assuming the minimum black hole mass is given by the mass function of
$5.8\pm0.5$\,M$_{\odot}$ \citep{Hynes:2003a}.  The true mass is likely
to be significantly larger than this, however, as the mass function is
only approached for high inclinations and there is no evidence that
\gx\ has a high inclination; indeed it has been argued that the
inclination is very low \citep{Wu:2001a}.  For actual black holes
masses of 10, 15, or 20\,M$_{\odot}$, distances of about 10, 12, or
14\,kpc respectively would be expected based on the method of
\citet{Maccarone:2003a}.  These are very reasonable if \gx\ lies
somewhat beyond the tangent point as we have argued.  If it is really
beyond 15\,kpc, however, then either the black hole mass is relatively
high, or the transition luminosity is rather high (but not implausibly
so).

\citet{Zdziarski:1998a} have instead tried to use the maximum observed
luminosity as a constraint on the distance.  We can reverse the
argument, however and examine the maximum luminosity for different
possible distances.  At 6\,kpc, their quoted maximum observed
luminosity corresponds to just 10\,\%\ of the Eddington limit of a
10\,M$_{\odot}$ black hole.  This seems rather low, and a higher black
hole mass would further exacerbate this.  In contrast, for a distance
of 15\,kpc and black hole masses of 10--20\,M$_{\odot}$, the Eddington
ratio is 0.3--0.6, which seems much more in line with other Galactic
black hole X-ray binaries.  More recent observations provide more
stringent constraints, yielding an 0.1--200\,keV flux of
$5.4\times10^{-8}$\,erg\,cm$^{-2}$\,s$^{-1}$ (extrapolated from a
3--100\,keV fit; Homan et al.\ in preparation).  For a 10\,M$_{\odot}$
black hole at distances of 6 and 15\,kpc respectively this corresponds
to 20\,\%\ or 110\,\%\ of the Eddington limit.  The latter could
readily be reconciled with a black hole somewhat more massive than
10\,M$_{\odot}$.  Thus the luminosity is consistent with a relatively
large distance given the implied high mass for the black hole.

\subsection{The `quiescent' luminosity}
The X-ray luminosity in the lowest observed state \citep{Gallo:2003a}
corresponds to $2\times10^{33}$\,erg\,s$^{-1}$ or
$1\times10^{34}$\,erg\,s$^{-1}$ for distances of 6 or 15\,kpc
respectively.  Most quiescent black holes have luminosities of between
$2\times10^{30}$ and $6\times10^{31}$\,erg\,s$^{-1}$
(\citealt{Garcia:2001a}; \citealt{Sutaria:2002a};
\citealt{Hameury:2003a}; \citealt{Tomsick:2003a};
\citealt{McClintock:2003a}).  The exception is V404~Cyg
($5\times10^{33}$\,erg\,s$^{-1}$; \citealt{Garcia:2001a}).  This is a
long period object (6.5\,day) and may be something of an anomaly among
quiescent black holes; indeed it exhibits dramatic X-ray variability
(\citealt{Wagner:1994a}; \citealt{Kong:2002a}; Hynes et al.\ in
preparation) which suggests that it may not be in a fully quiescent
state.  \gx\ appears to have a comparable or larger luminosity at
minimum light, so its lowest observed state may also not correspond to
the typical quiescent states of transient black hole binaries.
\subsection{The nature of the companion star}
A relatively active system is appealing in the light of the
non-detection of spectral features from the companion star
\citep{Shahbaz:2001a}.  Two explanations exist for this: either the
accretion flow is so luminous that it masks the companion star, or the
companion is an early-type star with no spectral features in the
region studied by \citet{Shahbaz:2001a}.  As noted above, a relatively
active accretion flow already favors a larger distance, and even
15\,kpc gives a plausible luminosity.  If we require an early-type
companion then its luminosity will be larger than for the typical K or
M companions of Galactic black holes, and so a large distance might
also required to avoid it being visible.

With no detection of the photospheric lines of the companion, its
nature is poorly constrained.  The companion star should be somewhat
evolved, as a 1.7\,day orbital period \citep{Hynes:2003a} implies a
mean density for the companion of only 0.06\,g\,cm$^{-3}$.  This is a
factor of ten below that obtained by \citet{Shahbaz:2001a} who assumed
a shorter orbital period.  Such a low density is reached on the main
sequence only for B0 stars; it is unlikely that the companion is such
a star, however, as it would then be expected to dominate the total
light at low luminosities, and also significantly affect the motion of
the black hole.  Hence the companion of \gx\ has a lower density than
a main-sequence star of the same temperature.  It is probably a
sub-giant, as late-type giants have even lower densities than we
derive for \gx.

For relatively early types, the companion mass is also significant and
should result in measurable orbital motion of the accretion disk.
\citet{Hynes:2003a} found no such evidence (only a weak modulation was
present, with the wrong phase-dependence) and suggested a mass ratio
of $q \la 0.08$.  This argues for a companion mass of 1.6\,M$_{\odot}$
or less unless the black hole mass is significantly larger than in
other objects ($\ga 20$\,M$_{\odot}$) and hence a spectral type of F
or later unless the companion is hotter than a main sequence star of
the same mass.  In support of these arguments, GRO\,J1655--40 has a
comparable orbital period (2.6\,d), and an F companion
(\citealt{Orosz:1997a}; \citealt{Shahbaz:1999a}).  Its mass ratio is
$0.42\pm0.03$ \citep{Shahbaz:2003a}, however, the disk lines move with
an amplitude of $76.2\pm7.5$\,km\,s$^{-1}$ \citep{Soria:1998a}, and
the companion totally dominates the optical light in quiescence and
remains visible even during outburst.  These properties are very
different to \gx, which evidently has a much less influential
companion.  G or K subgiants therefore seem more likely than earlier
types.

At our proposed distances there is little difficulty in accommodating
such companions.  We consider a number of examples.  For each we adopt
the main sequence mass corresponding to the spectral type, but adjust
the radius to fill the Roche lobe.  The size of the lobe is
constrained by the 1.7\,day orbital period, an assumed black hole mass
between 5.8 and 20\,M$_{\odot}$, and a mass ratio less than 0.1.  The
reddening is assumed to be between 0.8 and 1.6, as discussed above.
The upper limit is more pertinent, as it makes it easier to hide
luminous companions.  An F0 star (adopted mass 1.6\,M$_{\odot}$) could
only be accommodated if the black hole mass were near 20\,M$_{\odot}$;
otherwise we would expect detectable motion of the accretion disk.  At
6\,kpc such a star should have an $R$ magnitude of 17.0--19.5,
dependent on reddening.  This is brighter than the observed faintest
magnitude ($R=20.1$; \citealt{Shahbaz:2001a}), so is unlikely unless
the companion is significantly less massive (and hence has a smaller
Roche lobe).  At 15\,kpc, however, this could be as faint at $R=21.5$,
so could be masked by the accretion light.  For a G0 star (assuming
$M=1.1$\,M$_{\odot}$), we expect a range of 18.1--20.6 at 6\,kpc, or
20.1--22.6 at 15\,kpc, and either distance is acceptable, provided the
reddening is relatively large.  A K0 star ($M=0.8$\,M$_{\odot}$)
yields $R=19.1-22.7$ at 6\,kpc and 21.1--24.7 at 15\,kpc and poses no
difficulties.

These estimates are rather crude and dependent on some assumptions, but
are probably still useful as approximations.  To summarize, the
companion to \gx\ is likely to be a sub-giant, of spectral type G or
later if close, $\sim6$\,kpc, or possibly F if the
distance is much larger.  

%
\section{Conclusions}
\label{ConclusionSection}
We have analyzed \NaI\ and \CaII\ absorption along the line of sight
to \gx.  We find a rich absorption spectrum with a large velocity
dispersion.  Velocities are predominantly blue-shifted and span the
range of velocities expected for Galactic rotation along this
line-of-sight. The presence of multiple features with a significant
column density near the tangent point velocity suggests that \gx\ is
either at or beyond the tangent point, i.e.\ $d\ga 6$\,kpc.  The
presence of absorption at red-shifted velocities, together with the
systemic velocity of \gx\ itself, suggest the possibility that the
object could be at an even greater distance, on the far-side of the
Galaxy, $d\ga15$\,kpc.  Neither of these pieces of evidence are as
secure as that for a location beyond the tangent point, however.

Our analysis also indicates a total \NaI\ column density along the
line of sight of at least $3.5\times10^{13}$\,cm$^{-2}$, and this
suggests a reddening of at least $\ebv=0.5$, depending on assumed
ratios of \NaI\ to dust densities.  This is consistent with more
straightforward calibrations of the equivalent widths, which imply
$\ebv \ga 0.85$, but it could be considerably larger dependent on the
degree of saturation.  Hence even with well resolved observations of
the \nad\ lines it is impossible to derive a precise reddening from
these lines and published estimates should be viewed with caution.

We have examined the implied outburst and quiescent luminosities, and
the nature of the companion, in the context of these proposed
scenarios.  Neither a distance of 6\,kpc, nor 15\,kpc, presents
difficulties in these respects.  In fact the larger distance
simplifies matters somewhat, for example in explaining the
non-detection of the companion (likely a late-type sub-giant) in
quiescence.  Consequently we propose that a large distance,
$\sim15$\,kpc, should be given consideration.

%
\acknowledgments
We are grateful to Jeroen Homan, Jon Miller, and Seth Redfield for
productive discussions on this topic and to Tom Maccarone for helpful
comments on the manuscript.  RIH is supported by NASA through Hubble
Fellowship grant \#HF-01150.01-A awarded by STScI.  DS acknowledges a
Smithsonian Astrophysical Observatory Clay Fellowship.  JC
acknowledges support from the Spanish Ministry of Science and
Technology through the project AYA2002-03570.  This work uses
observations collected at ESO in Chile and the Anglo-Australian
Telescope.  We would particularly like to thank the ESO Director's
Office and VLT staff for a generous and efficiently executed award of
Director's Discretionary Time.  This work has also made use of the
NASA ADS Abstract Service and we are grateful to Craig Markwardt for
making the {\sc mpfit} IDL fitting routines publicly available.

\clearpage


\end{document}